\documentstyle[aps,preprint,epsfig]{revtex}

\begin{document}
\tightenlines \draft

\title{Oscillator strengths of dark charged excitons in the
\\ fractional quantum Hall regime}

\author{C.\ Sch\"uller, K.-B.\ Broocks, Ch.\ Heyn,
and D.\ Heitmann}

\address{Institut f\"ur Angewandte Physik und Zentrum
f\"ur Mikrostrukturforschung,\\ Universit\"at Hamburg,
Jungiusstra{\ss}e 11, D-20355 Hamburg, Germany}

\maketitle

\date{today}

\begin{abstract}
By direct absorption spectroscopy and comparison to
photoluminescence (PL), we investigate negatively charged excitons
in dilute two-dimensional electron systems at temperatures down to
$T=40$ mK in the regime of the fractional quantum Hall effect. At
very low temperatures, for filling factor $\nu < 1/3$, an
additional excitation appears in the PL spectrum, between the
well-known singlet and triplet excitons. The observation of a
similar excitation by PL was reported very recently [G.\ Yusa et
al., cond-mat0103561], and the excitation was assigned, in spite
of a PL intensity similar to that of the neutral exciton, to be
due to a dark triplet exciton. By comparing PL and direct
absorption spectra in optically thin samples at $T<100$ mK, we can
identify the new excitation indeed as a 'dark' mode, since we find
that the oscillator strength is much smaller than those of the
'bright' modes.
\end{abstract}

\pacs{71.30.+h 71.35.+z 73.20.Dx 78.66.Fd}

The fractional quantum Hall effect (FQHE), which is a direct
manifestation of many-particle interactions in a two-dimensional
electron system (2DES), has attracted great and sustaining
research interest since its discovery \cite{Stoermer}. The past
decade has shown that, in addition to transport experiments,
optical spectroscopies, especially photoluminescence (PL)
spectroscopy, can provide valuable further information about the
ground state properties of the systems. In fact, this method has
in recent years been applied to study an interactiong 2DES in the
regime of the FQHE
\cite{Turberfield,Goldberg,Kukushkin1,Kukushkin2}, where a variety
of anomalies in both, the energetic positions and the intensities
of the PL lines, were found. The striking difference in optical
experiments, compared to electron transport, is the presence of
photo-excited holes. Though it was previously shown that in
symmetric systems in the high magnetic field limit the
electron-electron and electron-hole interactions cancel exactly
due to a hidden symmetry \cite{MacDonald,Dzyubenko,Apalkov}, for
most of the real systems this symmetry is violated.

In order to systematically understand the interactions between
electrons and photo-excited holes in modulation-doped systems, it
seems natural to start with zero density of the 2DES and gradually
increase the number of free electrons. In fact, it was found that
in dilute 2DES the Coulomb attraction between electrons and holes
leads to the formation of negatively charged excitons, consisting
of two electrons and one hole \cite{Kheng,Finkelstein1,Shields1}.
It is now commonly accepted that at zero magnetic field the two
electrons form a spin singlet state (total spin ${\bf S}=0$). At
finite magnetic field, an additional bound state exists, where the
electrons form a spin triplet (${\bf S}=1$)
\cite{Shields2,Finkelstein2}. It was calculated that at high
magnetic fields the triplet exciton ($X_{t}^{-}$) forms the ground
state of the system. Magnetic fields of 30-40 T were predicted for
the singlet triplet crossing for {\em symmetric} quantum-well
systems \cite{Wojs,Palacios,Whittaker}. For some time,
experimental attempts to detect this singlet triplet crossing
failed. Instead, a saturation of the $X_{t}^{-}$ binding energy
was found at high fields \cite{Shields2,Finkelstein2,Kim,Hayne}.
This puzzle was recently solved by a theoretical work of A.\ Wojs
et al. \cite{Wojs1}. The authors calculated the spectrum of
charged excitons by exact diagonalization of finite size systems.
It was found \cite{Wojs1} that at high magnetic field {\em two
different} triplet excitons exist as bound states. One of these
excitons has a total angular momentum of ${\bf L}=0$, and is
therefore called the {\em bright} triplet ($X_{tb}^{-}$). The
second triplet exciton is called the {\em dark} triplet
($X_{td}^{-}$) because it can not decay radiatively (${\bf
L}=-1$). The calculations showed that the $X_{td}^{-}$ becomes the
ground state at high fields and the experimentally observed
saturation of the binding energy can be attributed to the bright
triplet. In a subsequent work, I.\ Szlurfarska et al.\ found that
in {\em asymmetric} quantum wells the singlet triplet crossing can
be at considerably lower fields \cite{Szlurfarska}. Very recently,
the observation of an excitation in PL experiments, which was
interpreted as a dark triplet exciton, was reported by two
different groups \cite{Munteanu1,Yusa}. The singlet triplet
crossing occured at $B=40$ T \cite{Munteanu1,Munteanu2} and $B=15$
T \cite{Yusa}. However, there seem to be some discrepancies
betweeen these reports, e.g., in Ref.\ \cite{Munteanu1} the dark
triplet is visible over the whole magnetic field range.

In this paper we report results which are concerning the PL
experiments consistent with the observation of G.\ Yusa et al.
\cite{Yusa}, i.e., (i) we observe a charged exciton at very low
temperatures, which is not present in the spectrum at $T=2$ K,
(ii) this exciton occurs for filling factor $\nu < 1/3$ only,
i.e., related to the FQHE regime, and, (iii) the PL intensity of
this excitation is of similar strength as that of the $X^0$.

At the first sight it is intriguing that a {\em dark} mode should
appear in the PL spectrum with almost equal strengths as the {\em
bright} modes, since its oscillator strength should by definition
be exactly zero. However, it is well known that the PL intensity
and linewidth might not reflect directly the oscillator strength
of a transition, which is proportional to $1/\tau_X$, where
$\tau_X$ is the lifetime of the excitation. The PL intensity is
proportional to $N_X/\tau_X$, i.e., it does also depend on the
number of excitons $N_X$. Due to relaxation and scattering
processes, which can take place prior to the recombination
process, the excitons at lower energies may have a larger
population than the excitons at higher energies. Therefore, in PL,
the factor $N_X$ may overcome a small oscillator strength
$1/\tau_X$ and lead to a significant PL signal. A direct
determination of the oscillator strength, however, can only come
from an absorption measurement. Therefore, we have prepared
optically thin samples. We were able to integrate external gates
to the thinned samples, which allow us to tune the carrier density
down to the dilute regime. This enabled us to measure, to the best
of our knowledge, for the first time, {\em direct} absorption of
charged excitons at millikelvin temperatures in a dilution
cryostat and compare it to PL. We were thus able to determine the
oscillator strengths of the observed excitations. We believe that
in our investigation we can clearly identify this charged exciton
as a dark exciton of the GaAs quantum well for two main reasons.
(i) In the optically thin samples no GaAs bulk material is left so
that we can definitely rule out any bulk-related effects. (ii) By
comparing PL {\em and} transmission at very low temperatures, we
find that the oscillator strength of the excitation labeled
$X_{td}^{-}$ is close to zero which unambiguously shows that this
is a dark mode.

The samples are one-sided modulation-doped GaAs-AlGaAs single
quantum wells, consisting of a 25 nm wide GaAs quantum well with a
88 nm AlGaAs spacer between the quantum well and the doped barrier
region on one side, and a 100 period AlGaAs/GaAs (10 nm / 3 nm per
period) superlattice on the other side. On top of the sample, a 10
nm thick titanium gate was deposited. The sample was glued upside
down on a glass substrate using an UV curing optical adhesive.
Subsequently, the sample was thinned from the back side by a
selective etching process \cite{LePore} down to the superlattice
to a total thickness of about 1.3 $\mu$m. By applying a negative
gate voltage, we can tune the density in the range between about
$1\times 10^{10}$ cm$^{-2}$ and $1\times 10^{11}$ cm$^{-2}$. PL
and absorption measurements were performed via glass fibers in a
$^3$He/$^4$He dilution cryostat at a base temperature of $T=40$ mK
and magnetic fields up to 16 T. A sensor at the sample position
indicated that during illumination the temperature directly at the
sample is about $T=100$ mK, while the base temperature is still
$T=40$ mK. Circularly polarized light was created directly inside
the mixing chamber and left and right circularly polarized spectra
were measured by ramping the magnet from positive to negative
magnetic fields. For the PL, the sample was excited by a
Ti:Sapphire laser at 750 nm. For the absorption measurement, a
white light source was used.

Figure 1 shows a comparison of left-circularly polarized PL
spectra, which were recorded at $T=2$ K (upper spectrum) in an
optical split-coil magnet, and at $T= 100$ mK (lower spectrum) in
the dilution cryostat. In this polarization configuration, at
$T=2$ K, the well-known lower Zeeman components of the singlet
exciton $X_{s}^{-}$ and the neutral exciton $X^0$, and the bright
triplet $X_{tb}^{-}$ with parallel spin orientation of the
electrons are visible. In the spectrum at $T=100$ mK a strong line
(labeled $X_{td}^{-}$) appears in between the $X_s^-$ and the
$X_{tb}^-$. We will argue below that this exciton is indeed a {\em
dark} triplet exciton. Due to small sample inhomogeneities, the
peaks in different measurement series do not always occur at
exactly the same absolute positions. Therefore, in Fig.\ 1, the
spectrum at $T=0.1$ K has been shifted by 0.16 meV so that the
positions of the $X^0$ is the same in both spectra. For both
spectra, the density is in the range $(2\pm 1) \times 10^{10}$
cm$^{-2}$. We have recorded extensive series of spectra for
different gate voltages in dependence on magnetic field.
Throughout the whole magnetic field range, the $X_{td}^-$ is not
detectable at all at $T=2$ K. For $T=100$ mK, our finding is that
the $X_{td}^-$ is present in the spectra for filling factors
$\nu<1/3$, which is consistent with the results reported in Ref.\
\cite{Yusa}. As examples, in Fig.\ 2, series of left-circularly
polarized spectra are displayed for two different gate voltages,
i.e., two different densities in the range of $10^{10}$ cm$^{-2}$.
For the higher electron density (Fig.\ 2a) one can see that the
neutral exciton $X^0$ is not visible in the displayed magnetic
field range and the $X_{td}^-$ splits from the bright triplet
$X_{tb}^-$ for $\nu<1/3$. At significantly lower density (Fig.\
2b), this splitting occurs at lower magnetic field ($\approx 2$ T
in Fig.\ 2b), and, at high fields, the neutral exciton $X^0$ is
visible. Consistent with the behavior predicted in Ref.\
\cite{Wojs1}, the binding energy of the $X_{td}^-$ increases with
magnetic field and the PL line approaches the $X_s^-$ at high
field (see Fig.\ 2b). However, we note here that up to the highest
field in our experiment (16 T) we did not observe the predicted
singlet triplet crossing. We assume that it will occur at somewhat
higher fields. According to the calculations in Ref.\
\cite{Szlurfarska}, the magnetic field value for the crossing
should depend critically on the electron hole separation $d$
perpendicular to the 2DES sheet. By comparing our experimentally
observed binding energies of the $X_{td}^-$ with calculations
(e.g.\ \cite{Palacios,Szlurfarska}), we can estimate for our
structures $d<0.5$ nm.

In the following we demonstrate that the observed $X_{td}^-$ is
indeed a {\em dark} mode. Figure 3 depicts absorption spectra in
dependence on magnetic field at $T=100$ mK for an electron density
of $n\approx 2\times 10^{10}$ cm$^{-2}$. One can clearly see
absorption due to the {\em bright} excitons $X_s^-$, $X_{tb}^-$,
and $X^0$. Remarkably, in Fig.\ 3 the neutral exciton $X^0$ has
the highest absorption, i.e., the largest oscillator strength,
which is proportional to $1/\tau_{X_0}$. As explained in the
introduction, this is not directly reflected in the PL experiments
(see Fig.\ 4), since the intensity of a PL line does via
$N_X/\tau_X$ also depend on the number of excitons $N_X$. From
Fig.\ 4 one can see that the $X_{td}^-$ does not appear in the
absorption spectrum, though it exhibits a well developed line in
the PL spectrum. From our signal-to-noise ratio we can estimate
that the $X_{td}^-$ has an oscillator strength which is at least
one order of magnitude smaller than the oscillator strength of
$X^0$. This can be taken as a clear signature that the $X_{td}^-$
is indeed a {\em dark} mode. According to Ref.\ \cite{Wojs1}, the
total angular momentum of this mode is ${\bf L}=-1$. Therefore, in
an idealized system, without scattering, it can not decay
radiatively. This means, its lifetime $\tau$ should go to infinity
and hence the oscillator strength $1/\tau$ to zero. This is
directly reflected in the absorption spectra in Fig.\ 3. On the
other hand, due to scattering processes prior to recombination, in
PL the selection rule ${\bf L}=0$ can be lifted. We believe that
such scattering-assisted recombination can be the reason for the
occurence of $X_{td}^-$ in the PL spectrum, though it is still
puzzling why its PL strength is almost equal to those of the {\em
bright} excitons. We think that our experimental observation,
i.e., that the high PL intensity of the dark exciton is not of
intrinsic origin, is also very important for further theoretical
treatments of this excitation. Furthermore, by our special sample
preparation, which does not leave any GaAs bulk material in the
sample, we can exclude impurity-related bulk excitons as a
possible origin for the $X_{td}^-$ line. (We note that for the
spectra compared in Fig.\ 4, the PL was excited by a Ti:Sapphire
laser and the absorption was measured using white light, which
required different aligning procedures. Therefore, we can not
exclude slight shifts in the absolute positions of the observed
lines due to sample inhomogeneities. This means that stokes shifts
between absorption and PL can not seriously be extracted from
these measurements.)

In conclusion, by comparing PL and absorption spectra at very low
temperatures, we could identify a dark triplet exciton $X_{td}^-$
by showing that, in spite of a relatively high PL intensity, its
oscillator strength is very small and justifies the assignment as
a {\em dark} mode. We find that the $X_{td}^-$ occurs at filling
factors $\nu<1/3$ only and is not visible at $T=2$ K.

We acknowledge stimulating discussions with Arkadiusz Wojs, Israel
Bar-Joseph, and Go Yusa. This work was supported by the Deutsche
Forschungsgemeinschaft via SFB 508, project SCHU1171/1 and a
Heisenberg grant (SCHU1171/2).

\begin{figure}
\caption{Comparison of left-circularly polarized PL spectra for
$T=2$ K and $T=0.1$ K. The electron density is in the range
$n\approx (2\pm 1) \times 10^{10}$ cm$^{-2}$. }\label{Fig1}
\end{figure}

\begin{figure}
\caption{Left-circularly polarized PL spectra for different
magnetic fields for $T=0.1$ K and an electron density of (a)
$n\approx 4.8 \times 10^{10}$ cm$^{-2}$, and, (b) $n\approx
1.4\times 10^{10}$ cm$^{-2}$.} \label{Fig2}
\end{figure}

\begin{figure}
\caption{Absorption spectra at $T=0.1$ K in dependence on magnetic
field. For clarity, the spectra are shifted horizontally and
vertically. The carrier density is about $2\times 10^{10}$
cm$^{-2}$. The inset shows a comparison of PL and absorption at
$B=9$ T. } \label{Fig3}
\end{figure}

\begin{figure}
\caption{Comparison of unpolarized PL and direct absorption
spectra at $B=9$ T and $T=0.1$ K. The small cusp at $E=1529.9$ meV
in the absorption spectrum in is due to noise, resulting from the
normalization procedure. This is also evident from Fig.\ 3. }
\label{Fig4}
\end{figure}

\newpage
\begin{figure}
\epsfig{file=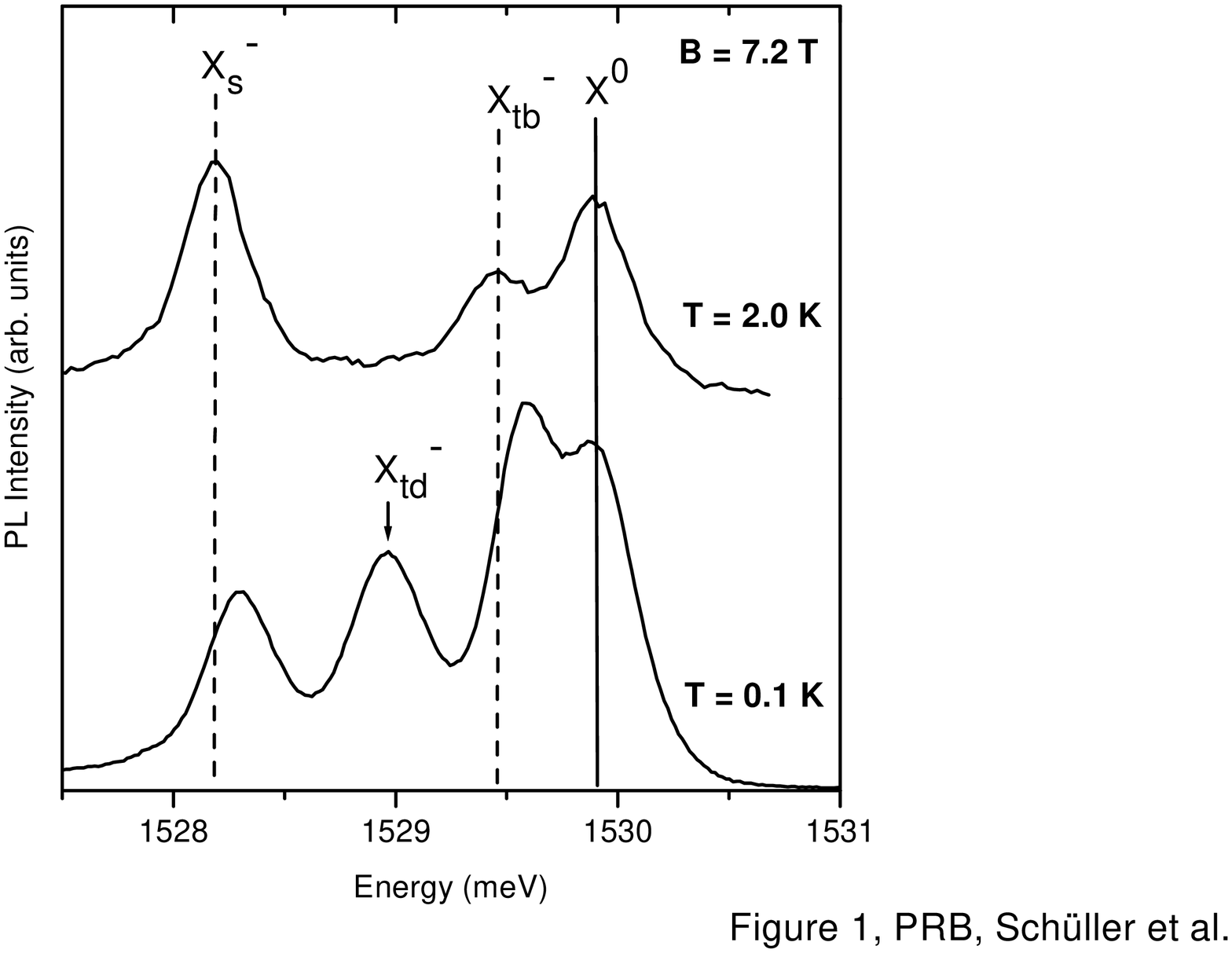, width=15cm}
\end{figure}
\newpage
\begin{figure}
\epsfig{file=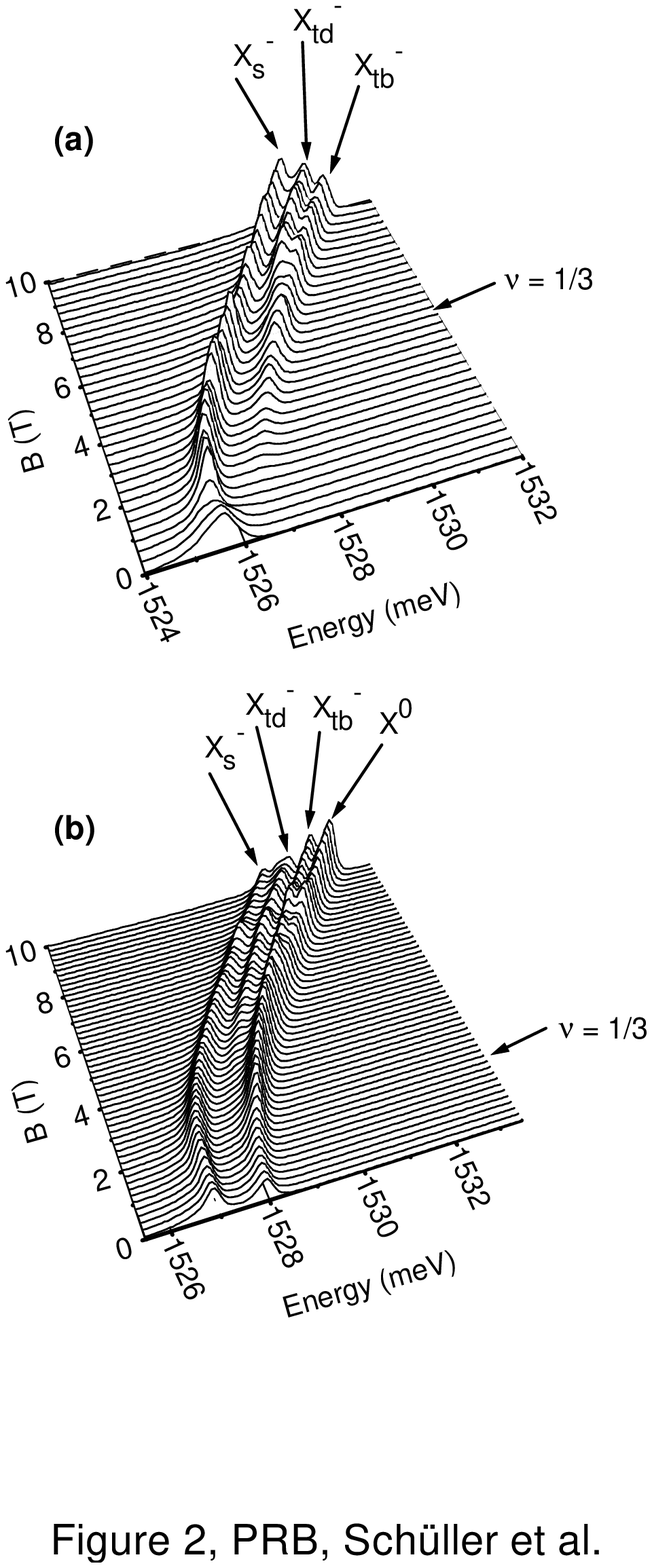, width=15cm}
\end{figure}
\newpage
\begin{figure}
\epsfig{file=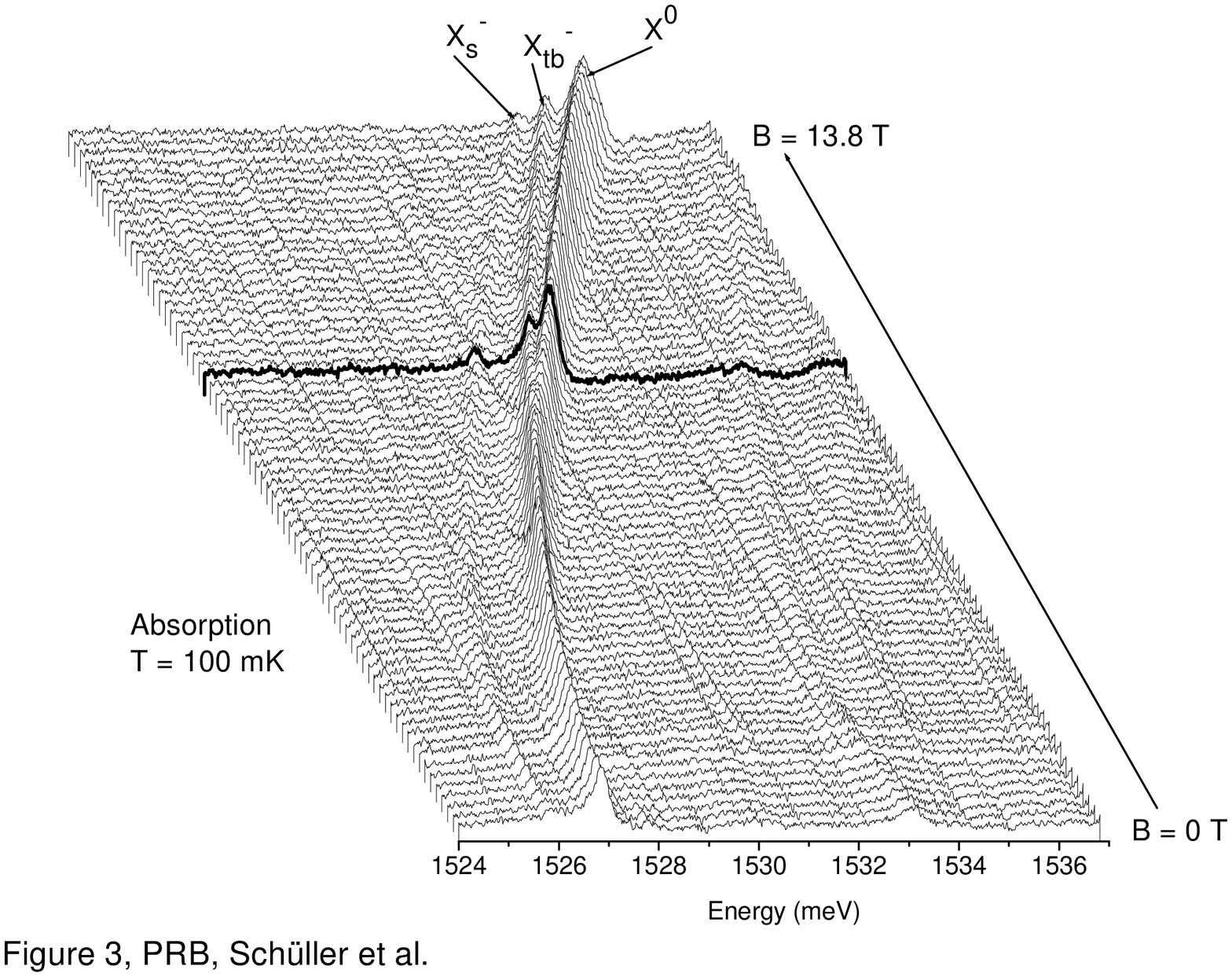, width=15cm}
\end{figure}
\newpage
\begin{figure}
\epsfig{file=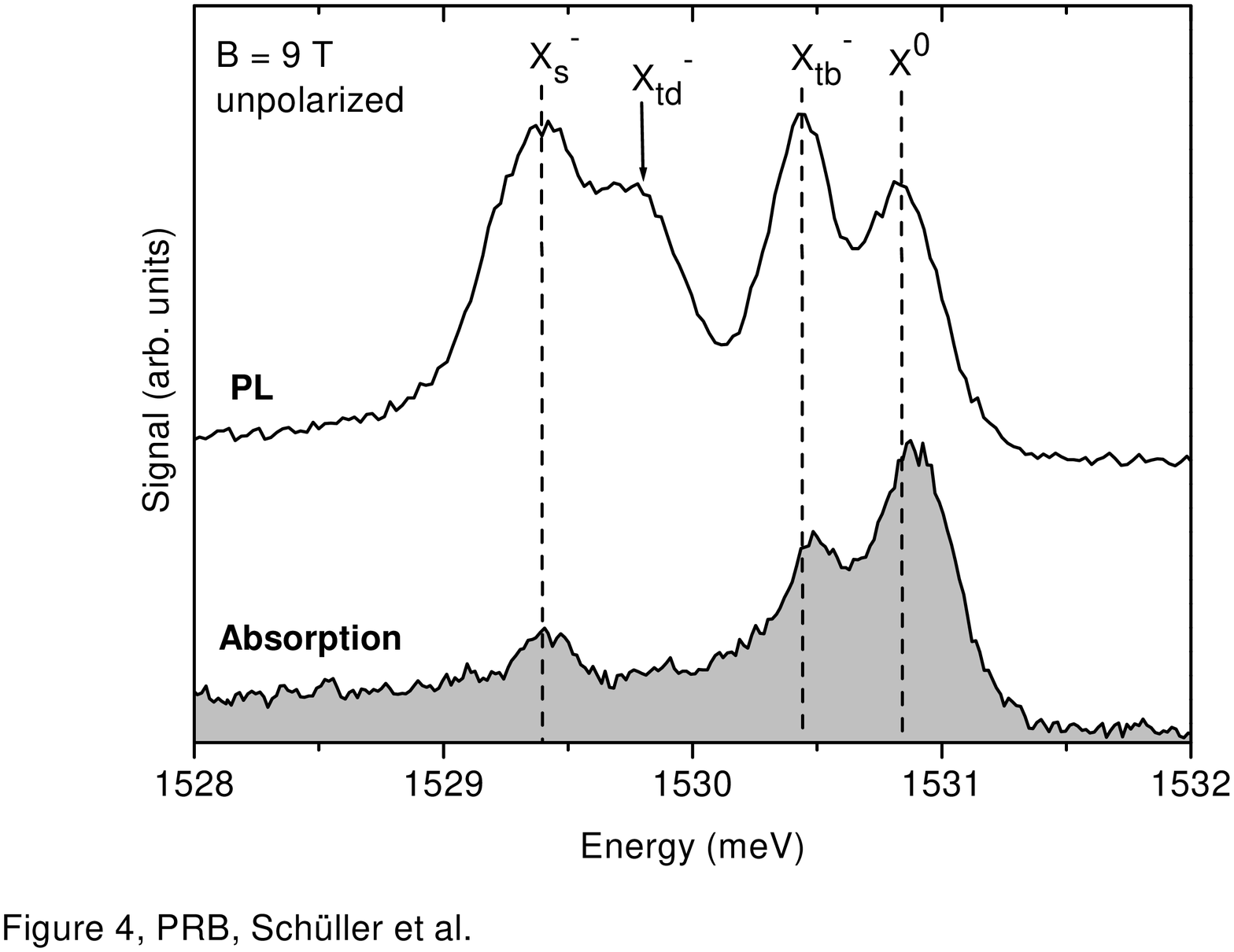, width=15cm}
\end{figure}

\end{document}